\documentclass[aps,prl,twocolumn,superscriptaddress,floatfix,longbibliography,citeautoscript]{revtex4-2}

\usepackage{array}[=2016-10-06]

\usepackage[normalem]{ulem}
\usepackage{enumitem}
\usepackage{xcolor}
\usepackage[colorlinks=true,urlcolor= blue,citecolor=blue,linkcolor= blue]{hyperref}
\usepackage{graphicx}
\usepackage{siunitx}
\usepackage{amsmath}
\usepackage{amssymb}
\usepackage{amsfonts}
\usepackage{bm}
\usepackage{booktabs}

\begin{document}

\title{Posture selection in active elastic filaments}

\author{Adam Pearl}
\affiliation{Department of Physics, Harvard University, Cambridge, Massachusetts 02138, USA}
\author{Ludwig A. Hoffmann}
\affiliation{School of Engineering \& Applied Sciences, Harvard University, Cambridge, Massachusetts 02138, USA}
\author{L. Mahadevan}
\email[]{lmahadev@g.harvard.edu}
\affiliation{Department of Physics, Harvard University, Cambridge, Massachusetts 02138, USA}
\affiliation{School of Engineering \& Applied Sciences, Harvard University, Cambridge, Massachusetts 02138, USA}
\affiliation{Department of Organismic and Evolutionary Biology, Harvard University, Cambridge, Massachusetts 02138, USA}

\date{\today}

\begin{abstract}
    Posture control in slender bodies such as snakes and eels arises from the interplay between passive deformation, active internal actuation, and task-level constraints. We formulate a general framework for the selection of stable postures in active elastic filaments subject to distributed forcing from gravity and fluid drag, by combining the constraints of mechanical equilibrium with optimal control theory. Our theory leads to a minimal description in terms of parameters governing the competition between hydrodynamic and gravitational loading, elasticity, and activity. We show that posture selection reflects a trade-off between control cost, function and dynamical stability, leading to the coexistence of distinct solution branches and abrupt transitions between them. Applying the theory to sessile eels in flow, we recover the experimentally observed transition from upright to reclining postures and predict scaling laws for body shape and exposed length. More generally, our results provide a unified perspective on how active filaments can regulate geometry to maintain function in external fields, with implications for biological and artificial systems.

\end{abstract}

\maketitle

\textit{Introduction}---Posture control in active flexible filamentous structures such as plant shoots, aquatic appendages, and soft robotic filaments interacting with external fields is a generic problem spanning biological and engineered systems. In all cases, the filaments or their assemblies strive to maintain configurations that emerge from the interplay of geometry, passive elasticity, internally generated actuation, and external loading from gravity, flow, or contact forces. They do so in the context of function defined in terms of posture that includes reaching, orienting, grasping and stable manipulation. A central question in the field is therefore how a particular posture is selected from the many physically admissible configurations.

In passive filaments, shape follows from a balance between elastic restoring forces and external loads, leading to well-characterized behaviors such as bending, buckling, and flow-induced reconfiguration~\cite{lighthill1960,batchelor2000,schouveiler2005,luhar2011,alben2002,argentina2005,gosselin2010,shelley2011}. In biological settings, filaments can actively regulate their own shape; e.g. in plants, posture control arises through growth and proprioceptive feedback in response to environmental cues~\cite{moulia2006,bastien2013,chelakkot2017,moulia2019,moulton2020,moulia2021}, while in animals, distributed muscular actuation and sensing enable flexible bodies to adapt dynamically to external forcing~\cite{tuthill2018,hoffmann2026}.

Here we formulate posture selection as a problem in the optimal control of active filaments under distributed forcing: the system must balance external loads to maintain a functional shape while minimizing a cost associated with deformation and actuation. Because active systems must complete tasks under dynamically varying conditions, we then analyze the dynamic stability of these equilibrium postures. Finally, we apply our framework to published experiments on sessile eels in a flow and explain how multiple steady configurations can coexist. We show that posture selection reflects a trade-off between energetic cost and dynamical stability, leading to bifurcations and abrupt transitions between distinct shapes.
\begin{figure}[b]
    \centering
    \includegraphics[width=\columnwidth]{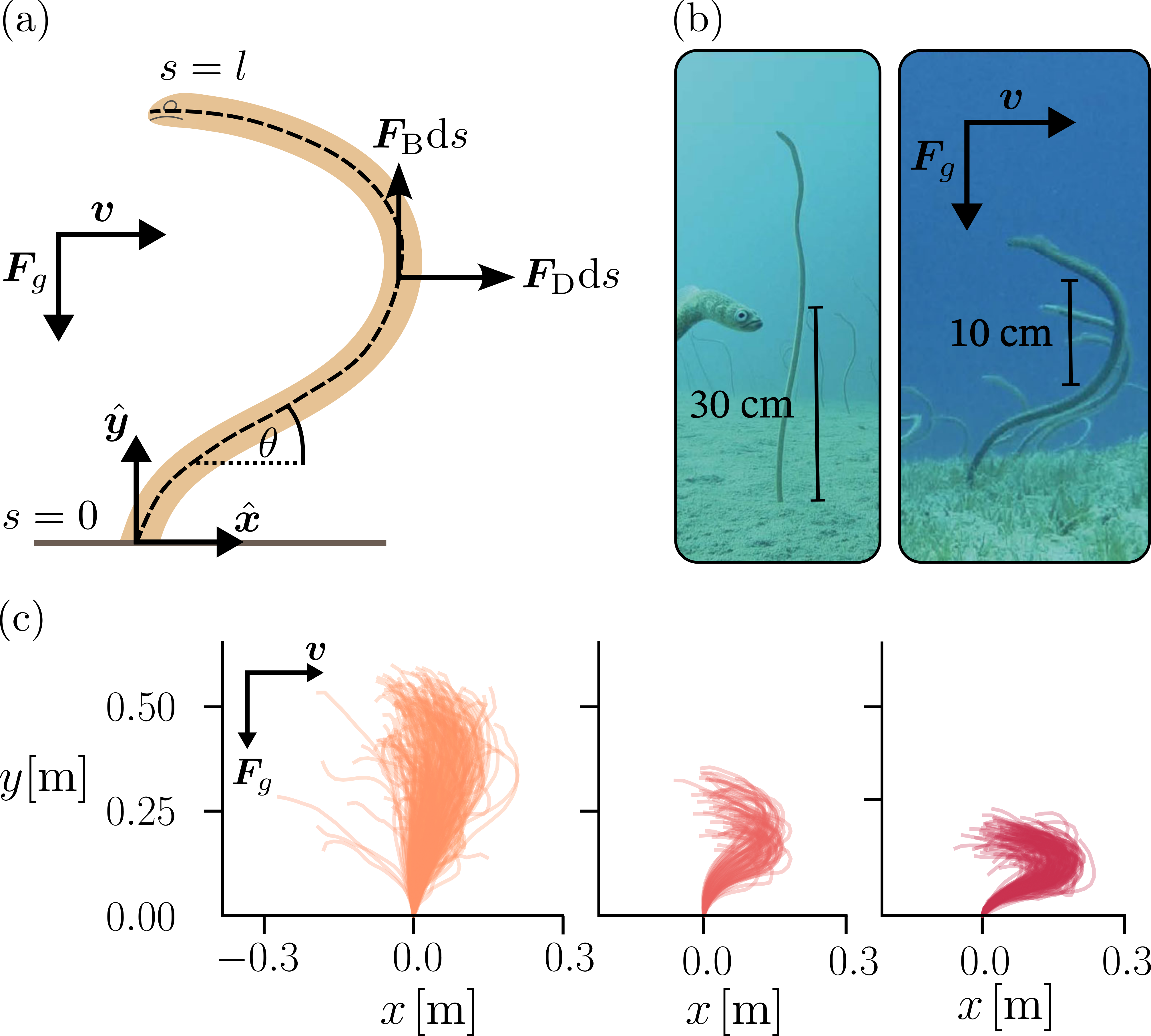}
    \caption{\textbf{Garden eels in external current.} (a) Sketch of an eel together with the midline (dashed) along which the coordinate $s \in [0,l]$ is defined. The relevant forces described in the main text are indicated. (b) Colony of garden eel species \textit{Gorgasia sillneri} in the Red Sea under a weak current (left, $4\, \si{ \cm\per\second}$) and strong current (right, $26\, \si{\cm\per\second}$). Approximate directions of current velocity $\bm{v}$ and gravitational force $\bm{F}_g$ are indicated. (c) Projection of 3D eel posture into the plane spanned by $\bm{v}$ and $\bm{F}_{g}$ for increasing current speeds of $3.1\, \si{\cm\per\second}$ (left), $16.6\, \si{\cm\per\second}$ (center), and $29.9 \, \si{\cm\per\second}$ (right). Reproduced from Ref.~\cite{khrizman2018}, see SM for details.}
    \label{fig:Model}
\end{figure}
\textit{Model}--- The geometry of the active elastic filament subject to distributed external forcing can be described in terms of its centerline $\bm{r}(s)$, parameterized by arc length $s \in [0,l]$, with unit tangent $\hat{\bm{t}} = \partial_s \bm{r}$ and normal $\hat{\bm{n}}$ (see Fig.~\ref{fig:Model}(a)). Restricting attention to planar inextensible deformations (a good assumption for snakes, eels and many robotic manipulators that have an endoskeleton), the configuration is specified by the tangent angle $\theta(s)$ via $\hat{\bm{t}} = (\cos\theta, \sin\theta)$. In the quasi-static limit, force balance reduces to $\partial_s \bm{N}(s) + \bm{F}_{\rm{ext}}(s) = 0$, where $\bm{N}$ is the internal force resultant and $\bm{F}_{\rm{ext}}$ the external force per unit length~\footnote{Projecting onto the normal direction defines the shear force $Q = \bm{N}\cdot \hat{\bm{n}}$, which satisfies ${\rm d}Q/{\rm d}s = \bm{F}_{\rm{ext}} \cdot \hat{\bm{n}}$.}.

The internal moment $M$ can be decomposed additively into passive and active contributions, $M = M_{\rm p} + M_{\rm a}$~\footnote{The additivity emerges in the asymptotic limit under the assumption of slender geometry.}. Here, $M_{\rm p} = B_{\rm p} \kappa$, where $B_{\rm p}$ is the passive bending stiffness and $\kappa = \partial_s \theta$ the local curvature. The active moment $M_{\rm a}(s)$ represents distributed actuation. Then, combining force and moment balance, introducing the dimensionless arc length $\sigma = s/l$, and rescaling the active moment as $\tilde{M}_{\rm a} = M_{\rm a} l / B_{\rm p}$, yields the governing equation for the shape
\begin{equation}
\frac{{\rm d}^3 \theta}{{\rm d}\sigma^3} + \frac{{\rm d}^2 \tilde{M}_{\rm a}}{{\rm d}\sigma^2}
= \mathcal{F}(\theta;\Lambda_i) \;.
\label{eq:general}
\end{equation}
Here, $\mathcal{F}$ represents the known external forcing, while the dimensionless parameters $\Lambda_i$ quantify its strength relative to elasticity. Within this framework, posture selection and control are associated with determining the functional form of $M_{\rm a}$, which naturally requires some additional information. 

The active moment may be described either by postulating a local feedback law or by solving a global optimal control problem~\cite{alvarado2026}. In the latter case, the functional form of $M_{\rm a}$ is obtained by minimizing a cost functional $J(\kappa,M_{\rm a})$, comprising a running cost $\int_{0}^{1} j_{\rm R}(\kappa,M_{\rm a})(\sigma)\, {\rm d}\sigma$, typically associated with actuation, and a terminal cost $J_{\rm T}(\kappa,M_{\rm a})|_{\sigma=1}$, associated with a desired end-point $\sigma=1$, with their precise forms determined by the application. The minimization is subject to Eq.~\eqref{eq:general}, the boundary conditions, and potential constraints on $\kappa$ or $M_{\rm a}$, and will result in a shape and actuation that globally minimize $J$ given the imposed constraints. Unlike conventional time-dependent optimal control, here the optimal trajectory represents the filament posture (see End Matter for a complete formulation).

Before considering the general framework for posture selection and control, we first consider the case of sessile organisms in fluid environments, which provide a clear realization of this problem. Garden eels anchor themselves in burrows and feed on prey transported by currents~\cite{fricke1970,khrizman2018,ishikawa2022,khrizman2024}; see Fig.~\ref{fig:Model}(b). Experiments~\cite{khrizman2018} show that they adopt distinct postures depending on flow velocity: upright and mobile in weak currents, and bent into a reclining configuration aligned with the flow in stronger currents; see Figs.~\ref{fig:Model}(b),(c). These postural changes suggest a strategy that reduces drag while maintaining foraging access to the flow, and are often accompanied by adjustments in exposed length, reflecting a trade-off between feeding efficiency and mechanical stability.

To make this idea concrete, we consider a filament in a uniform flow $\bm{v} = v \hat{\bm{x}}$, subject to a drag force per unit length 
$\bm{F}_{\rm{D}} = -\rho C_{\rm{D}} R (\bm{v} \cdot \hat{\bm{n}})^2 \hat{\bm{n}}$ and a buoyancy force per unit length  $\bm{F}_B = \Delta \rho g \pi R^2 \hat{\bm{y}}$. Here, $C_{\rm{D}}$ is the drag coefficient, $R$ the radius of a cross-section of the eel, $\rho$ the density of water, $\Delta \rho = \rho - \rho_{\text{eel}}$ the difference in density between eel and water, and $g$ the gravitational acceleration. Projecting these onto the normal direction of the filament and substituting into the governing Eq.~\eqref{eq:general} yields
\begin{subequations}
\label{eq:master}
\begin{equation}
\frac{{\rm d}^3 \theta}{{\rm d}\sigma^3}
- \Lambda_{\rm B} \cos\theta
+ \Lambda_{\rm D} \sin^2\theta
+ \frac{{\rm d}^2 \tilde{M}_{\rm a}}{{\rm d}\sigma^2}
= 0\;,
\end{equation}
where the dimensionless parameters
\begin{equation}
\Lambda_{\rm B} = \frac{l^3 \Delta \rho g \pi R^2}{B_{\rm p}}, 
\qquad
\Lambda_{\rm D} = \frac{l^3 \rho C_{\rm{D}} R v^2}{B_{\rm p}}
\end{equation}
\end{subequations}
measure the relative importance of buoyancy and drag compared to elastic bending. A summary of the parameters and their estimated values is provided in Tab.~S1. We note that in the limit $\Lambda_{\rm D} \to 0$, Eq.~\eqref{eq:master} reduces to the case of a filament subject solely to gravity, used recently to model snake postures~\cite{hoffmann2026} (see End Matter for further details).

To complete the formulation of the problem, we impose boundary conditions such that the filament is vertical at the base ($\theta(0) = \pi/2$) and force free at the tip ($\theta''(1)= 0$). In the passive limit $M_{\rm a}=0$, with a vanishing moment at the tip, $\theta'(1)=0$, Eq.~\eqref{eq:master} reduces to a problem that has been studied extensively~\cite{alben2002,argentina2005,gosselin2010,luhar2011,shelley2011} (see SM Sec.~S2 for a summary of these studies). For an active filament, we must specify the active moment and the boundary conditions that are task dependent, and below we consider two different approaches. 

\textit{Local proprioceptive feedback}---The ability to actively control and sense one's own shape and dynamically react to it (proprioceptive feedback) is known to be crucial in posture control~\cite{bastien2013,bastien2015,chelakkot2017,tuthill2018,moulia2019,moulia2021,hoffmann2026}. A simple model of such activity is to set $M_{\rm a} = B_{\rm a}\kappa$, with $B_{\rm a}$ a constant, i.e. the active moment due to musculature is proportional to the local curvature, which has been used to describe a variety of plant and animal postures~\cite{bastien2013,bastien2015,moulia2019,moulia2021,hoffmann2026}. Substituting this expression for $M_{\rm a}$ into Eq.~\eqref{eq:master} yields that activity leads to a rescaling of the bending stiffness, $B_{\rm p} \to B_{\rm p} + B_{\rm a}$: muscle activation effectively stiffens the body. Accordingly, we define the rescaled parameters $\bar{\Lambda}_i \equiv B_{\rm p}\Lambda_i/(B_{\rm a} + B_{\rm p})$ (see Tab.~S1 for estimates for biological values and SM Sec.~S3 for a discussion of the effects of buoyancy). In what follows we set $\Lambda_{\rm B} = 0$, as drag forces are dominant.

\begin{figure}
    \centering
    \includegraphics[width=3.3in]{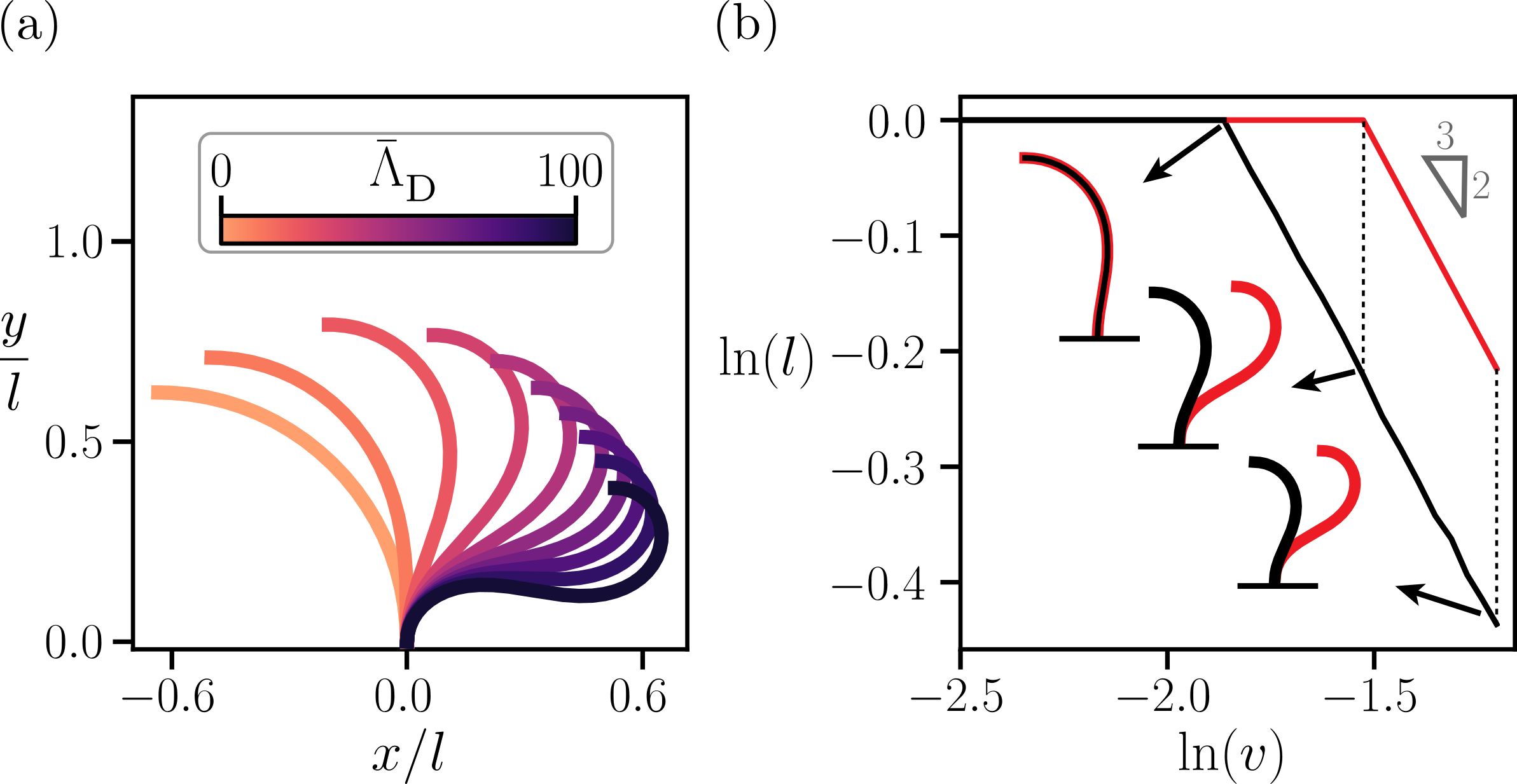}
    \caption{\textbf{Posture under proportional feedback.} 
    (a) Shape obtained from Eq.~\eqref{eq:master} for different values of the rescaled drag $\bar{\Lambda}_{\rm D}$, with $M_{\rm a} = B_{\rm a}\kappa$ and a forward-facing boundary condition. 
    (b) Exposed length $l$ that maximizes either head height (black curve) or total exposed length (red curve) under a fixed drag constraint, $\bar{\Lambda}_{\rm D}\int_0^1 \sin^2\theta {\rm d}\;\sigma \leq 25$, shown as a function of speed $v$ for an eel of total length $L= 1~\si{\meter}$. Representative postures for both cases are shown for three velocities. Beyond the inflection points, both curves scale as $l \sim v^{-2/3}$.}
    \label{fig:active}
\end{figure}

Solving Eq.~\eqref{eq:master} numerically for different values of $\bar{\Lambda}_{\rm D}$ and with $\theta(\sigma = 1) = \pi$, we obtain the postures shown in Fig.~\ref{fig:active}(a). The eel leans forward for small drag, but reclines backwards, approaching a characteristic ``question-mark'' shape, as $\bar{\Lambda}_{\rm D}$ is increased. This change in posture significantly reduces the drag forces acting on the eel, since the projected length perpendicular to the current is reduced, and is qualitatively similar to the experimentally observed postures (Fig.~\ref{fig:Model}(b)). 

In addition to changing posture, drag can be reduced by decreasing the exposed length $l$; however, larger $l$ increases the foraging volume, and greater head height provides access to higher plankton flux because the flow is slower near the seabed~\cite{khrizman2018,genin2024,khrizman2024}. To examine the trade-off between drag reduction and foraging performance, we treat $l \in [0,L]$ as an additional degree of freedom. For a prescribed maximum drag force above which the eel will likely dislodge, and a given fluid speed $v$, we ask how to maximize \textit{i}) total exposed length and \textit{ii}) head height. Results for one choice of maximum drag are shown in Fig.~\ref{fig:active}(b). As $v$ increases, the eel changes posture as before while remaining fully extended, $l=L$, until the drag reaches the prescribed limit. Thereafter, posture remains fixed while the exposed length decreases as $l \sim v^{-2/3}$ such that $\bar{\Lambda}_{\rm D} \sim \text{const}$ (see SM for a derivation). This immediately implies that $\bar{\Lambda}_{\rm D} \sim v^2 l^3/B_{\rm a}$, i.e. both increasing the active stiffness and reducing the exposed length decreases the rescaled drag. It is useful to contrast this with the passive case, where the same ``forward-facing'' solutions are in principle allowed (albeit at the same drag only for much smaller current speeds), if the moment balance condition at the tip is actively enforced (see SM Sec.~S2). We address this point in the following section.

\begin{figure*}
  \centering
  \includegraphics[width=7in]{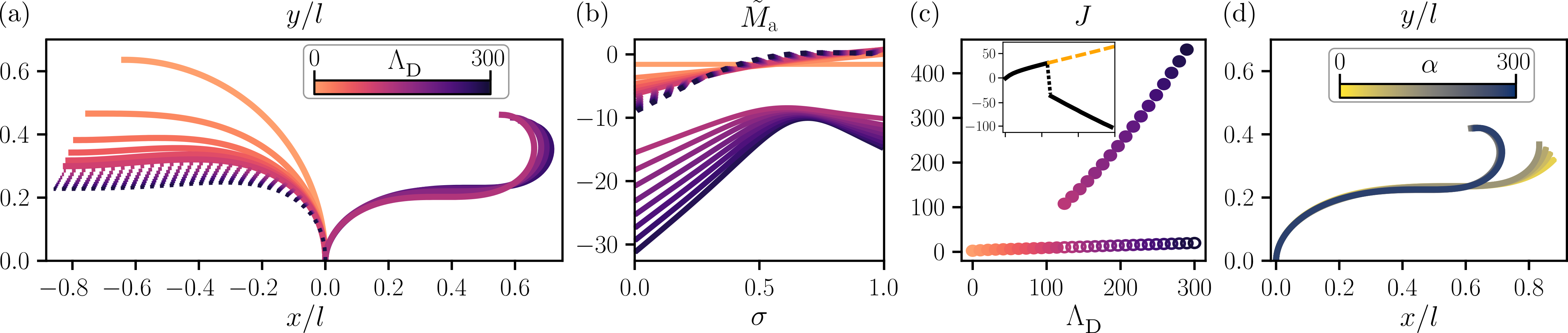}
  \caption{\textbf{Optimal posture.} 
  (a) Optimal postures obtained by solving Eq.~\eqref{eq:EelOptimalControl} for different values of the drag parameter $\Lambda_{\rm D}$. When two solutions exist for a given $\Lambda_{\rm D}$ (see main text), the forward-leaning and reclining postures are shown as dashed and solid curves, respectively.
  (b) Corresponding profiles of the active moment. 
  (c) Cost $J$ (Eq.~\eqref{eq:EelCost}), with solid and open circles corresponding to the solid and dashed shapes in (a), respectively. The inset shows the real part of the eigenvalues of Eq.~\eqref{eq:stability} for the postures in (a), with the dashed yellow (solid black) curve corresponding to the dashed (solid) solutions. The dotted portion of the black curve corresponds to the jump.
  (d) Optimal postures for different weights $\alpha$ at fixed $\Lambda_{\rm D} = 300$.
  For all plots we use $\mathcal{K} = 28$, $\mathcal{U} = 600$, $\mathcal{M} = \infty$, $\epsilon = 0.01$.}
  \label{fig:Control}
\end{figure*}

\textit{Global control}---In contrast with local proprioceptive feedback, we can also use a global control formulation to determine a functional form for $M_{\rm a}$. Then, we seek the filament configuration and actuation field that satisfy the governing equation~\eqref{eq:master} while minimizing a cost functional reflecting the energetic cost of deformation and control.

Introducing the dimensionless state vector $\bm{x}(\sigma) = \big(x, y, \theta, \tilde{\kappa}, \tilde{\kappa}', \tilde{M}_{\rm a}, \tilde{M}_{\rm a}'\big)$, where $\tilde{\kappa} = \theta'(\sigma)$, and defining the control $u = \tilde{M}_{\rm a}''(\sigma)$, Eq.~\eqref{eq:master} can be written as a first-order system $\bm{x}'(\sigma) = \bm{f}(\bm{x}, u)$, with
\begin{equation*}
\bm{f} = \big(\cos\theta, \sin\theta, \tilde{\kappa}, \tilde{\kappa}', \Lambda_{\rm B} \cos\theta- \Lambda_{\rm D} \sin^2\theta - u, \tilde{M}_{\rm a}', u\big).
\end{equation*}
Next, we define the cost functional
\begin{subequations}
\begin{equation}
J[u] = \int_0^1 \left( \tilde{M}_{\rm a}^2 + \epsilon u^2 \right) {\rm d}\sigma
+ \alpha \big[\theta(1) - \theta^*\big]^2 \;,
\label{eq:EelCost}
\end{equation}
which is the sum of the spatial cost of actuation ($\tilde{M}_{\rm a}^2$) and its rapid spatial variation ($\epsilon u^2$, with $\epsilon \ll 1$), and a terminal cost that enforces a preferred tip orientation $\theta^*$, with the parameter $\alpha$ controlling the relative importance of the terminal constraint. At the base, we impose an anchoring condition, $\theta(0) = \pi/2$, while at the tip we additionally impose moment- and force-free conditions,
\begin{equation}
\tilde{\kappa}(1) = -\tilde{M}_{\rm a}(1), \qquad
\tilde{\kappa}'(1) = -\tilde{M}_{\rm a}'(1) \;.
\label{eq:EelBCs}
\end{equation}
To restrict solutions to physically admissible regimes, we impose bounds on curvature, actuation, and control,
\begin{equation}
|\tilde{\kappa}| \le \mathcal{K}, \qquad
|\tilde{M}_{\rm a}| \le \mathcal{M}, \qquad
|u| \le \mathcal{U}.
\label{eq:EelConstraints}
\end{equation}
\label{eq:EelOptimalControl}
\end{subequations}
The optimal control problem is thus to minimize $J[u]$ given by Eq.~\eqref{eq:EelCost} subject to the state dynamics~\eqref{eq:master} and boundary conditions~\eqref{eq:EelBCs} and constraints~\eqref{eq:EelConstraints}. This formulation is equivalent to a spatial optimal control problem in which arc length plays the role of time, and the active moment acts as a distributed control that shapes the filament. We solve this optimal control problem numerically using \texttt{CasADi}~\cite{andersson2019} (see SM for details).

First, we consider the limit $\alpha \to \infty$ and set $\theta(1) = \pi$ to directly compare with the proprioceptive-feedback results above. For different values of $\Lambda_{\rm D}$, the results for posture and active moment are shown in Fig.~\ref{fig:Control}(a),~(b). For small $\Lambda_{\rm D}$, we again obtain a forward-leaning solution. However, upon increasing the drag, the eel continues to extend further forward (dashed curves in Fig.~\ref{fig:Control}(a)) and never transitions to a reclining posture. In contrast, a reclining solution is found if we start from a large value of $\Lambda_{\rm D}$. This posture is maintained as the drag is decreased, until a critical value of $\Lambda_{\rm D}$ is reached, when the solution jumps discontinuously to a forward-leaning configuration (solid curves in Fig.~\ref{fig:Control}(a)). This bifurcation is clearly reflected in the cost (Fig.~\ref{fig:Control}(c)). The reclining posture is associated with a much larger cost and muscular activation (Fig.~\ref{fig:Control}(b),~(c)), indicating that the reclining branch corresponds only to a local minimum of the cost landscape. 

Why, then, would eels adopt the posture associated with a larger cost?  We hypothesize that it is associated with an increased stability: it must be possible to maintain the posture under perturbations of shape and flow, and the forward-leaning shape may require substantial effort to correct and recover from perturbations, as small changes in the angle significantly increase the drag force. 

Assuming a separation of time scales, i.e. the reaction time is long compared with the time scale on which an unstable perturbation grows, we consider the linear stability of a given solution $\{\theta_0(\sigma),u_0(\sigma)\}$ of the optimal-control problem, while keeping $\tilde{M}_{\rm a}$ fixed. A linearized dynamical version of Eq.~\eqref{eq:master} for small perturbations $\delta \theta(\sigma)$ relative to the optimal control solution then yields:
\begin{align}
\label{eq:stability}
\partial_t \delta \theta \approx -\delta \theta ''' -\Lambda_{\rm D}\delta\theta\sin[2\theta_0] + \mathcal{O}(\delta\theta^2).
\end{align}
To determine the stability of the posture, we numerically compute the eigenvalues of the linear operator on the right-hand side for the different shapes in Fig.~\ref{fig:Control}(a). The results are shown in the inset of Fig.~\ref{fig:Control}(c). Postures in the forward-leaning regime have eigenvalues with positive real parts that increase with $\Lambda_{\rm D}$. In contrast, postures in the reclining regime have eigenvalues with negative real parts that become more negative as $\Lambda_{\rm D}$ increases. Therefore, the reclining solution is indeed more stable, providing a rationale for adopting it despite its higher energetic cost.

We now consider the role of the terminal cost associated with pointing into the flow, by varying $\alpha$. Increasing $\alpha$ from zero for fixed $\Lambda_{\rm D}$ with $\theta^* = \pi$ causes the angle $\theta(1)$ to increase, so that the eel's head bends progressively toward the flow; see Fig.~\ref{fig:Control}(d). It is found that initially $\theta(1)$ increases smoothly, reflecting a trade-off between placing the head at its preferred angle and minimizing muscular activation. Above a critical value of $\alpha$, however, the posture jumps to a configuration with $\theta(1) \approx \pi$; see Fig.~S4. As in Fig.~\ref{fig:Control}(a), we observe a subcritical bifurcation in the energy (see SM Fig.~S4(c)) with an unstable solution at intermediate head angles.

Experimental data~\cite{khrizman2018} on eel posture show that at low flow velocities, eels remain mostly upright and exhibit substantial out-of-plane motion (Fig.~S6). At higher velocities, out-of-plane motion is suppressed and the observed postures collapse to characteristic shapes in agreement with our predictions (see SM Sec.~6 for a quantitative comparison and a discussion of length change as well as results for a different eel species~\cite{ishikawa2022}).

\textit{Discussion}---Starting from a model of an active elastic filament, we formulated a general framework for posture control using two approaches (proprioceptive feedback and optimal control theory). As a concrete example, we applied the model to sessile eels exposed to background currents of varying intensity. The competition between energetic cost and stability gives rise to multiple coexisting steady configurations and abrupt transitions between them. At sufficiently high current speeds, our theoretical predictions agree well with experimental observations. 

\begin{figure}
 \centering
 \includegraphics[width=\columnwidth]{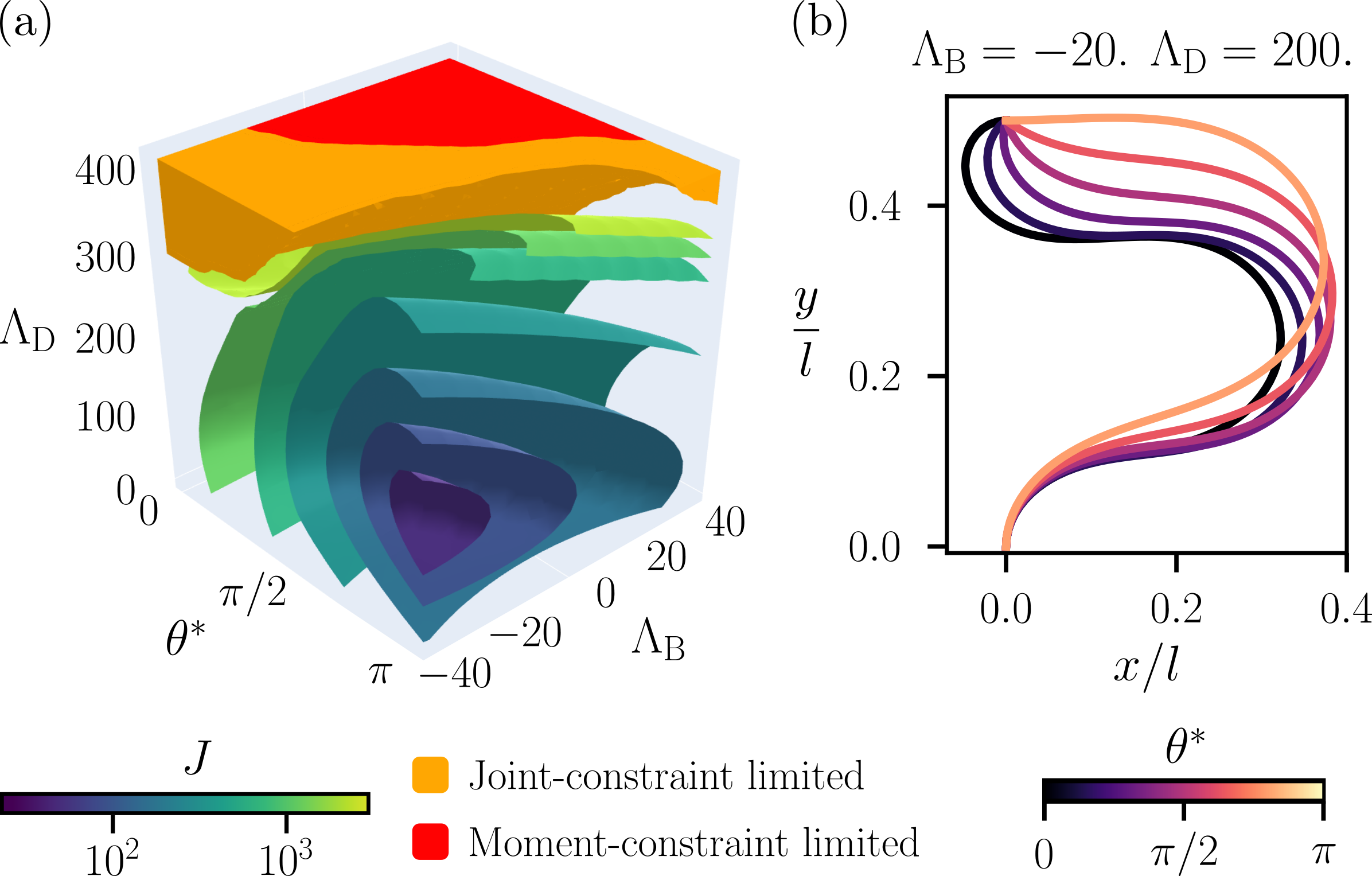}
 \caption{\textbf{Feasibility diagram.} (a) Solution feasibility of the general control problem (cf. Eq.~\eqref{eq:EelOptimalControl}) for an active elastic filament subject to hydrodynamic drag $\Lambda_{\rm{D}}$ and buoyancy $\Lambda_{\rm{B}}$ with head orientation $\theta^*$. We choose boundary conditions $\theta(0) = \pi/2$, $x(1)/l=0$, and $y(1)/l = 0.5$. At each point, we solve Eq.~\eqref{eq:EelOptimalControl} under four choices of constraints $\mathcal{B} \equiv \{\mathcal{M},\mathcal{K}\}$: unconstrained ($\mathcal{B}_1=\{\infty,\infty\}$), curvature constrained ($\mathcal{B}_2=\{\infty,\mathcal{K}_{\rm b}\}$), active-moment constrained ($\mathcal{B}_3=\{\mathcal{M}_{\rm b},\infty\}$), and jointly constrained ($\mathcal{B}_4=\{\mathcal{M}_{\rm b},\mathcal{K}_{\rm b}\}$). Cost isosurfaces are shown in regions that are feasible under any of the above. Orange region is feasible under $\mathcal{B}_{1-3}$ while the red region is only feasible under $\mathcal{B}_{1,2}$. Here, $\mathcal{M}_{\rm b} = 35$ and $\mathcal{K}_{\rm b} = 13$. (b) Example of shapes for varying head orientation, while keeping $\Lambda_{\rm B} = -20$ and $\Lambda_{\rm D} = 200$ fixed.}
 \label{fig:drag_scaling}
\end{figure}

Our general framework for posture control allows us to construct a ``cost space'' that quantifies how expensive it is to achieve a posture, or whether it is even possible given constraints on activity, external forces and boundary conditions. In Fig.~\ref{fig:drag_scaling}(a), we show an example in which gravitational load, drag, and head angle $\theta^*$ span the parameter space of passive external forces, with relevance to a wide range of applications (see End Matter), while tip angle and tip position are associated with function. Example shapes are shown in Fig.~\ref{fig:drag_scaling}(b). We find that the cost isosurfaces foliating parameter space are highly non-trivial and curved, resulting in non-monotonic behavior as parameters are varied (see SM Fig.~S7). In particular, varying buoyancy produces the smallest change in cost, while changes are greatest along the $\theta^*$-axis. In addition, certain regions of parameter space are accessible when either curvature or active moment is constrained, but not when both are jointly constrained (orange region in Fig.~\ref{fig:drag_scaling}(a)). This reflects the possibility that, if only one field is constrained, the system can compensate for one field reaching its limit by increasing the range of the other. Finally, embedded in this region is a smaller region that is always inaccessible when active moment is constrained (red region in Fig.~\ref{fig:drag_scaling}(a)), where this ``compensation'' fails for sufficiently large $\Lambda_{\rm D}$. A second example with a different tip position is shown in SM Fig.~S8; the behavior is qualitatively similar.

Natural extensions of our work include coupling posture selection explicitly to feeding performance and burrowing stability, and testing some of our quantitative predictions in artificial systems such as bio-inspired soft robots.

\section{Acknowledgments}
We thank Alexandra Khrizman for sharing the data of Ref.~\cite{khrizman2018} with us, and acknowledge the NWO Rubicon grant (LAH), the Simons Foundation (LM), and the Henri Seydoux Fund (LM) for partial financial support.

\bibliographystyle{apsrev4-2}
\bibliography{Biblio.bib}

\section{End Matter}
\setlength{\tabcolsep}{6pt}
\renewcommand{\arraystretch}{1.3}
\begin{table*}
    \centering
    \begin{tabular}{p{0.16\textwidth} p{0.165\textwidth} p{0.215\textwidth} p{0.23\textwidth} p{0.1\textwidth}}
    \toprule
        System 
        & External load(s) [$\mathcal{F}$] 
        & Actuation mechanism [$M_{\rm a}$]
        & Function [$\bm{x}_{\rm term}$] 
        & References \\

    \midrule
        Garden eel 
        & Fluid drag 
        & Distributed axial muscles 
        & Orient head for drift-feeding 
        & \cite{khrizman2018,khrizman2024}, here \\

        Arboreal snake 
        & Gravity
        & Distributed axial muscles
        & Span gaps or lift body
        & \cite{jayne2007,byrnes2012,jayne2020,hoffmann2026} \\

        Plant shoot 
        & Gravity \& wind drag 
        & Differential growth 
        & Set growth/shoot orientation 
        & \cite{bastien2013,chelakkot2017,moulia2019} \\

        Elephant trunk 
        & Gravity
        & Muscular hydrostat
        & Reach or grasp
        & \cite{dagenais2021,kaczmarski2024,agabiti2026} \\

        Octopus arm 
        & Fluid drag
        & Muscular hydrostat
        & Reach or grasp
        & \cite{gutfreund1996,sumbre2005,flash2023,shih2023} \\

        Soft robotic filament 
        & Gravity \& fluid drag
        & Engineered actuation
        & Reach, grasp, or manipulate
        & \cite{trivedi2008,rus2015,laschi2016,liu2026} \\

    \bottomrule
    \end{tabular}
    \caption{
    Representative systems that can be described within the one-dimensional posture-control framework. The source of actuation, the dominant external loads, and the functional terminal objective differ across systems, but in each case the posture can be formulated as a mechanically admissible shape selected by actuation costs, endpoint requirements, and system-specific constraints. Some representative references are given. Engineered systems may use a range of actuation mechanisms, including pneumatic, hydraulic, or tendon-driven actuation.
    }
    \label{tab:tab1}
\end{table*}

Here we outline how the framework developed above for garden-eel posture can be extended to posture control in a broader class of slender, flexible bodies with local activation. We consider a quasi-one-dimensional body whose centerline shape is described by the curvature field $\kappa(s)$. The body is assumed to be passively elastic, while active control enters through a distributed active moment $M_{\rm a}(s)$. Together, passive elasticity, active moment, and external loading determine the state of the filament which we assume is given by force balance as written in Eq.~\eqref{eq:general}.

The boundary conditions specify how the filament is positioned and oriented at the two ends, $\sigma \equiv s/l=0$ and $\sigma=1$, and are associated with the mode of anchoring and the orientation of the free end, itself a characteristic of the function, e.g. feeding, grasping, manipulation, etc.

A mechanically admissible posture must satisfy the governing equation while meeting these boundary conditions, but does not, in general, select a unique shape. We therefore assume that the realized posture is selected by an optimality principle: among all admissible configurations, the system chooses the one that minimizes a prescribed cost functional that represents total actuation, mechanical work, deviation from a desired terminal state, or a combination of these quantities. With this interpretation, posture control can be formulated in the general form

\begin{subequations}
\label{eq:control}
\begin{align}
    \min_{u \in \mathcal{U}} J&(u)
    = \int_{0}^{1} j_{\rm R}(\bm{x}, u)\, {\rm d}\sigma + J_{\rm T}(\bm{x}(1)) \label{eq:Cost} \\
    \text{s.t.} \quad
    &\bm{x}'(\sigma) = \bm{f}(\bm{x}(\sigma), u(\sigma)) \label{eq:EOM} \\
    &\bm{g}\left(\bm{x}(0),\,\bm{x}(1)\right) = 0
    \label{eq:BC1} \\
    &\bm{h}(\bm{x}(\sigma),u(\sigma)) \le 0 \;. \label{eq:Constraints}
\end{align}
\end{subequations}
Here, $u=\tilde{M}_{\rm a}''$ is the control variable, an \emph{a priori} unspecified function drawn from an admissible set $\mathcal{U}$. This choice is convenient because it casts the governing equation as a first-order system; the physically relevant actuation remains the active moment $\tilde{M}_{\rm a}$ itself. The control variable is varied so as to minimize the cost functional $J(u)$ subject to the governing equations, Eq.~\eqref{eq:EOM}, where the state vector is $\bm{x}(\sigma) = (x, y, \theta, \tilde{\kappa}, \tilde{\kappa}', \tilde{M}_{\rm a}, \tilde{M}_{\rm a}')$, and, from Eq.~\eqref{eq:general}
\begin{equation}
\bm{f} =\left(\cos\theta,\sin\theta,\tilde{\kappa},\tilde{\kappa}',\mathcal{F}(\theta;\Lambda_i)-u,\tilde{M}_{\rm a}',u\right).
\end{equation}
The running cost $j_{\rm R}$ measures the cost of actuation along the body. In the examples considered here, we use costs based on the active moment, for example $j_{\rm R} = |\tilde{M}_{\rm a}|^2$,
which penalizes the squared actuation amplitude, or $j_{\rm R} = \tilde{M}_{\rm a}\tilde{\kappa}$, which provides a work-like measure of bending actuation.  Terminal requirements may be imposed either as hard constraints, by prescribing some components of $\bm{x}(1)$, or as soft constraints through the terminal cost $J_{\rm T}$. The latter formulation is useful when the terminal state represents a preferred functional target rather than an exact constraint. All hard boundary conditions are collected in the compact form $\bm{g}(\bm{x}(0),\bm{x}(1))=0$.

Finally, the path constraint $\bm{h}(\bm{x},u)\le 0$ represents anatomical, material, or design limitations. For instance, finite joint mobility or tissue extensibility may impose a curvature bound, $|\tilde{\kappa}|\le \mathcal{K}$, whereas finite muscle strength or actuator capacity may impose $|\tilde{M}_{\rm a}|\le \mathcal{M}$. Without them, the mathematical optimum may require curvatures or active moments that are physically unrealistic.

This formulation can be adapted to a range of posture-control problems, as summarized in Tab.~\ref{tab:tab1}, including garden eels and snakes bridging vertical or horizontal gaps~\cite{jayne2007,byrnes2012,jayne2020,hoffmann2026} where muscular actuation is the control, growing plant shoots subject to proprioceptive and gravitropic responses~\cite{delangre2008,bastien2013,hamant2016,chelakkot2017,moulia2019} where growth fields are the control, elephant trunks and octopus arms~\cite{gutfreund1996,sumbre2005,dagenais2021,shih2023,flash2023,kaczmarski2024,agabiti2026}, where the terminal state of a functional end-effector dominates the control, and soft robotic filaments designed for grasping, manipulation, or locomotion~\cite{hannan2003,trivedi2008,rus2015,laschi2016,pettersen2017,shih2023,kaczmarski2024,alessi2024,liu2026}, where actuation and terminal costs are often balanced.
\end{document}

\title{Supplementary Material: Posture selection in active elastic filaments}

\author{Adam Pearl}
\affiliation{Department of Physics, Harvard University, Cambridge, Massachusetts 02138, USA}
\author{Ludwig A. Hoffmann}
\affiliation{School of Engineering \& Applied Sciences, Harvard University, Cambridge, Massachusetts 02138, USA}
\author{L. Mahadevan}
\email[]{lmahadev@g.harvard.edu}
\affiliation{Department of Physics, Harvard University, Cambridge, Massachusetts 02138, USA}
\affiliation{School of Engineering \& Applied Sciences, Harvard University, Cambridge, Massachusetts 02138, USA}
\affiliation{Department of Organismic and Evolutionary Biology, Harvard University, Cambridge, Massachusetts 02138, USA}

\date{\today}
\maketitle

\section{Parameter values}
\renewcommand{\arraystretch}{1.2}
The parameters used in this work, along with their units and estimated biological values where available, are listed in Tab.~\ref{tab:params_table}. The ranges for $\Lambda_{\rm{B}}$ and $\Lambda_{\rm{D}}$ are calculated from Eq.~(2b) using the parameter ranges given in Tab.~\ref{tab:params_table}.

\begin{table*}[h!]
\centering
\begin{tabularx}{\textwidth}{l>{\centering\arraybackslash}X>{\centering\arraybackslash}X>{\centering\arraybackslash}X}
\toprule
\textbf{Parameter name} & \textbf{Symbol} & \textbf{Value / Range} & \textbf{Units} \\
\midrule
Eel length & $L$ & $0.24 \le L \le 0.68$~\cite{khrizman2018} & \si{\meter} \\
Exposed eel length & $l$ &  $0 \le l \le L$& \si{\meter} \\
Water density & $\rho$ & 1028 ~\cite{khrizman2018}& \si{\kilogram\per\meter\cubed} \\
Eel density & $\rho_{\text{eel}}$ & 1000 ~\cite{long1998}& \si{\kilogram\per\meter\cubed} \\
Gravitational acceleration & $g$ & 9.81 & \si{\meter\per\second\squared} \\
Eel cross-sectional radius & $R$ & 0.005 ~\cite{khrizman2018} & \si{\meter} \\
Current velocity & $v$ & $0 \le v \le 0.3$ ~\cite{khrizman2018} & \si{\meter\per\second} \\
Passive elasticity & $B_p$ & $1.5\times10^{-4}$ \cite{long1998} & \si{\newton\meter\squared} \\
\midrule
\multicolumn{4}{c}{\textbf{Dimensionless parameters}} \\
\midrule
Drag coefficient & $C_{\rm D}$ & $0.5 \le C_{\rm D} \le 1$ ~\cite{khrizman2018}& 1\\
Drag parameter & $\Lambda_{\rm{D}} = l^3C_D R\rho v^2/B_p$ & $0 \le \Lambda_{\rm{D}} \le 411$ & 1 \\
Buoyancy parameter & $\Lambda_{\rm{B}} = l^3\Delta\rho g \pi R^2/B_p $ & $2 \le \Lambda_{\rm{B}} \le 45$ & 1\\
Curvature constraint & $ \mathcal{K}$ & \text{Unknown} & 1\\
Active moment constraint & $ \mathcal{M}$ & \text{Unknown} & 1\\
Control constraint & $\mathcal{U}$ & \text{Unknown} & 1\\
\bottomrule
\end{tabularx}
\caption{\textbf{Parameters and their estimated values.} Dimensional parameters (top) and dimensionless parameters (bottom), together with their respective units and definitions. Biological values or ranges are estimated from the cited references.}
\label{tab:params_table}
\end{table*}

\newpage
\section{Passive shape}
\begin{figure}[b]
    \centering
    \includegraphics[width=\linewidth]{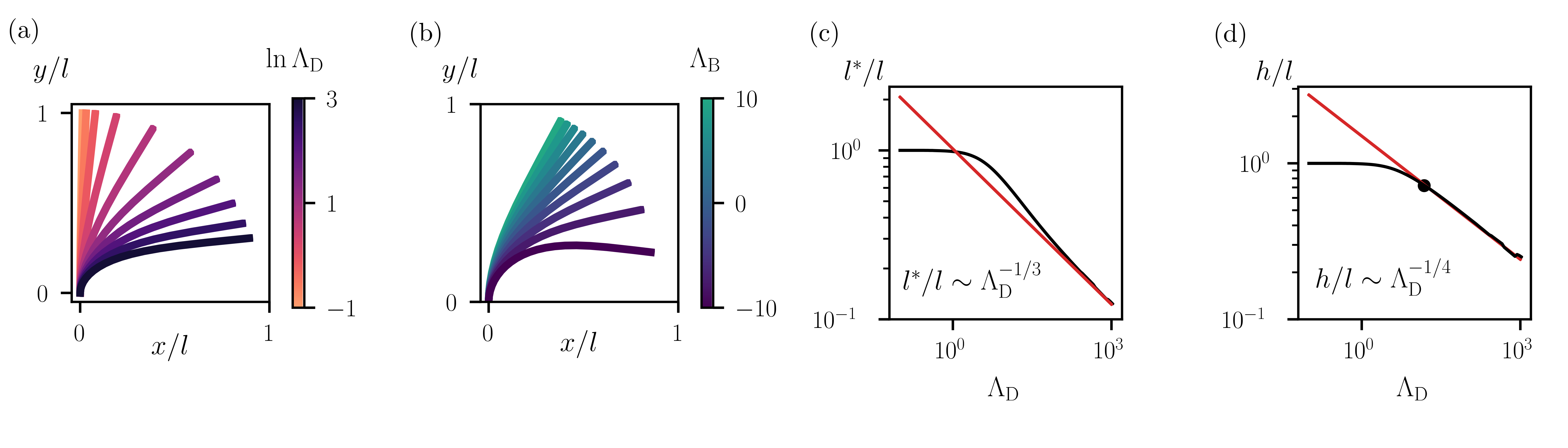}
    \caption{\textbf{Passive postures}. Solutions to Eq.~\eqref{eq:SImaster} subject to force-free and moment-free boundary conditions at the tip. (a) Filament shape as a function of the drag parameter on a logarithmic scale, with $\Lambda_{\rm{B}} = 0$. (b) Filament shape as a function of the buoyancy parameter, with $\Lambda_{\rm{D}} = 10$. (c) Log-log scaling of the effective length, defined in Eq.~\eqref{eq:effective_length}, with $\Lambda_{\rm{D}}$. The black curve corresponds to the numerical solution, while the red curve shows the scaling $l^{*} / l \sim \Lambda_{\rm{D}}^{-1/3}$. (d) Log-log scaling of the non-dimensional head height with the drag parameter. The black curve corresponds to the numerical solution, while the red curve shows the scaling $h / l \sim \Lambda_{\rm{D}}^{-1/4}$.}  
    \label{fig:passive}
\end{figure}
Here we revisit Eq.~(2) of the main text for a passive filament in the absence of active moments, i.e., $M_{\rm a} = 0$, for which the governing equation reduces to
\begin{equation}
    \label{eq:SImaster}
    \frac{{\rm d}^3\theta(\sigma)}{{\rm d}\sigma^3} - \Lambda_{\rm B} \cos \theta(\sigma) + \Lambda_{\rm D}\sin^2\theta(\sigma) = 0 \;.
\end{equation}
We impose boundary conditions corresponding to an upright filament at the base, $\theta(0) = \pi/2$, and a moment- and force-free tip, $\left.{\rm d}\theta/{\rm d}\sigma\right|_{\sigma = 1} = 0$ and $\left.{\rm d}^2\theta/{\rm d}\sigma^2\right|_{\sigma = 1} = 0$, respectively. We solve Eq.~\eqref{eq:SImaster} numerically and present the resulting postures in Fig.~\ref{fig:passive}(a), (b) for different values of $\Lambda_{\rm{D}}$ and $\Lambda_{\rm{B}}$, respectively.

To quantify the drag force experienced by a filament in a given posture, we define the effective length $l^*$ of a filament as the exposed length of a filament (for which $\theta = \pi/2$ uniformly along its length) that would experience the same drag force~\cite{luhar2011}. For this, note that the total drag force experienced by a filament of exposed length $l$ in a posture $\theta(\sigma)$ is
\begin{equation*}
    \int_0^l F_{\rm D} \,{\rm d}s
    = R l \rho C_{\rm D} v^2 \int_0^1 \sin^2\theta \,{\rm d}\sigma \;.
\end{equation*}
In contrast, an upright filament of length $l^*$, with $\theta(\sigma) = \pi/2$, experiences a total drag force $\rho C_{\mathrm{D}} R v^2 l^{*}$. Equating these two expressions gives
\begin{equation}
   l^*/l = \int_0^1 \sin^2\theta \,{\rm d}\sigma \;,
\label{eq:effective_length}
\end{equation}
which depends only on the filament shape and is independent of any constants.

Since the non-dimensional drag force experienced by a filament is given by $\Lambda_{\mathrm{D}}\int_0^1\sin^2\theta \,{\rm d}\sigma = \Lambda_{\mathrm{D}} l^*/l$, the non-dimensional effective length corresponds to the drag force per unit $\Lambda_{\mathrm{D}}$. In Fig.~\ref{fig:passive}(c), we show the numerically obtained effective length $l^*/l$ as a function of drag. For large values of $\Lambda_{\mathrm{D}}$, we find that $l^*/l \sim \Lambda_{\mathrm{D}}^{-1/3}$, and therefore the drag force scales as $F_{\mathrm{D}} \sim \Lambda_{\mathrm{D}}^{2/3}$. To understand this scaling, note that at large current velocities, the restoring force due to stiffness is localized near the base. The bending force scales as $\sim B_{\mathrm{p}}(1/l^{*})^2$. Balancing this restoring force with the drag force gives $B_{\mathrm{p}}(1/l^{*})^2 \propto \rho C_{\mathrm{D}} R v^2 l^{*}$, and therefore $l^{*} \sim v^{-2/3}$ and $l^{*}/l \sim \Lambda_{\mathrm{D}}^{-1/3}$. This scaling is consistent with that reported in Ref.~\cite{luhar2011}. Finally, in Fig.~\ref{fig:passive}(d), we plot the rescaled head height $h/l$, where $h = y(\sigma = 1)$, as a function of drag. For large drag, we find that $h/l \sim \Lambda_{\mathrm{D}}^{-1/4}$.

\section{Effect of buoyancy}
\begin{figure}[b]
    \centering
    \includegraphics[width=\linewidth]{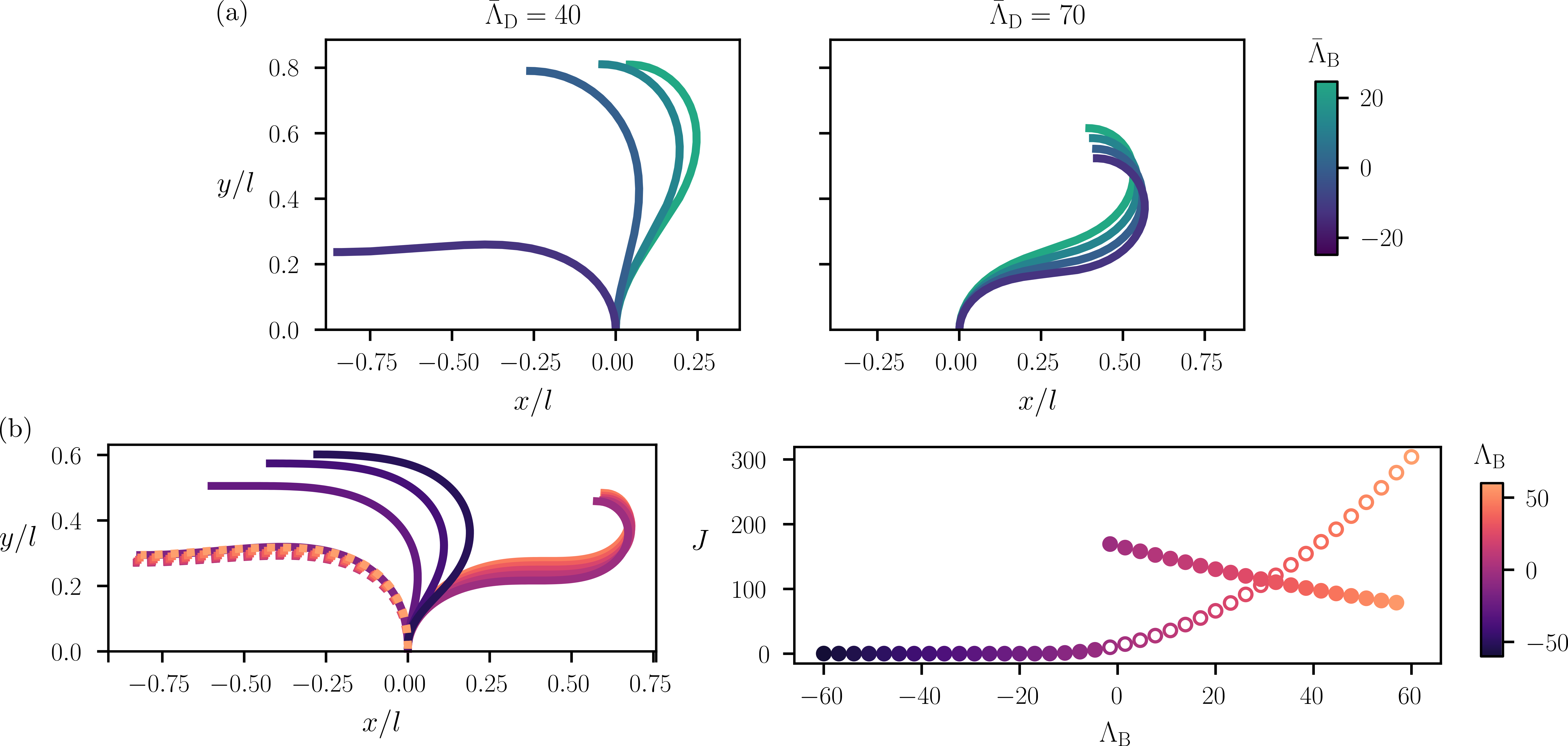}
    \caption{\textbf{Postures for varying buoyancy}. 
    (a) Solutions to Eq.~(2) for an internally actuated filament with force-free, forward-facing boundary conditions at the tip. Shapes under varying buoyancy at a fixed drag of $\bar{\Lambda}_{\rm{D}} = 40$ (left) and $\bar{\Lambda}_{\rm{D}} = 70$ (right). Under a negatively buoyant net force, the eels adopt an equilibrium posture closer to the seabed for small buoyancy.
    (b) Optimal postures obtained by solving Eq.~(3) for different values of the buoyancy parameter $\Lambda_{\rm{B}}$ at fixed drag, $\Lambda_{\rm{D}} = 160$. When two solutions exist for a given $\Lambda_{\rm{B}}$ (see main text) the forward-leaning and reclining postures are shown as dashed and solid curves, respectively. The cost $J$ associated with each solution is shown in the right panel, with open circles corresponding to the dashed curves in the left panel. We used $\mathcal{K} = 28$, $\mathcal{M} = \infty$, $\mathcal{U} = \infty$, and $\epsilon = 0.01$.}
    \label{fig:buoyancy_SI} 
\end{figure}

In the main text, we assumed for simplicity that buoyancy is negligible and set $\Lambda_{\rm{B}} = 0$ in order to investigate how posture changes with the drag parameter, $\Lambda_{\rm{D}}$. However, in Tab.~\ref{tab:params_table}, we estimate a range of $2 \le \Lambda_{\rm{B}} \le 45$, suggesting that the eel is in fact positively buoyant. Here, we repeat the analysis performed in the main text for both the local proprioceptive feedback and global control approaches, now varying $\Lambda_{\rm{B}}$ while keeping $\Lambda_{\rm{D}}$ fixed.

To model local proprioceptive feedback, we set $M_{\rm a} = B_{\rm a}\kappa$, as described in the main text. We solve the resulting equation, Eq.~(2) of the main text, numerically with boundary conditions $\theta(0)=\pi/2$, $\theta''(1) = 0$, and $\theta(\sigma = 1) = \pi$ for a range of $\bar{\Lambda}_{\rm B}$ while keeping $\bar{\Lambda}_{\rm D}$ fixed. The resulting postures for two fixed values of $\bar{\Lambda}_{\rm D}$ are shown in Fig.~\ref{fig:buoyancy_SI}(a). We observe that the eel adopts a lower equilibrium position under negative buoyancy, and shifts upwards as buoyancy is increased, both for smaller velocities (left panel) and larger velocities (right panel). In the left panel, we see some evidence of a phase transition between lunging and reclining regimes, which will become more evident in the following optimal control approach.

To investigate the effect of non-zero buoyancy in the optimal control approach, we solve Eq.~(3) of the main text at fixed drag, $\Lambda_{\rm{D}} = 160$, for a range of $\Lambda_{\rm{B}}$ values. The results are shown in Fig.~\ref{fig:buoyancy_SI}(b). Similar to what was found in Fig.~3(a), we observe a bifurcation in the postures, with one lunging posture and one reclining posture. This bifurcation occurs as $\Lambda_{\rm{B}}$ crosses zero, as clearly reflected in the cost function. When we start from a negative value of $\Lambda_{\rm{B}}$ and iteratively increase it, the eel adopts the reclining posture, whereas when we start from a positive value of $\Lambda_{\rm{B}}$ and iteratively decrease it, the eel adopts the lunging posture. However, the costs associated with these postures differ significantly from those in Fig.~3 of the main text. The cost associated with lunging postures increases monotonically with $\Lambda_{\rm{B}}$, whereas the cost associated with reclining postures decreases monotonically with $\Lambda_{\rm{B}}$ (see the right panel of Fig.~\ref{fig:buoyancy_SI}(b)). As a result, there is a critical value of $\Lambda_{\rm{B}}$ above which reclining postures are more energetically favorable than lunging postures. This is in contrast to the results of Fig. 3 of the main text, where reclining postures are always less energetically favorable than lunging postures. For $\Lambda_{\rm{D}} = 160$, this critical value occurs at $\Lambda_{\rm{B}} \approx 30$, which lies within the range of plausible values reported in Tab.~\ref{tab:params_table}.

We provide the following intuitive explanation for this effect. Under strong positive buoyancy in the lunging posture, eels must counter both the drag force acting on the upright portion of their lower bodies and the upward buoyancy force acting on the horizontal portion of their upper bodies. Both forces contribute to a positive torque. Whereas $J$ scales approximately linearly with $\Lambda_{\rm{D}}$ for lunging postures in Fig.~3(c), here $J$ scales superlinearly because of the additional adversarial buoyancy force. In contrast, in the reclining regime, the buoyancy force provides a negative torque that counteracts the positive torque from drag, thereby reducing the energetic cost. If eels are indeed positively buoyant, this effect could provide an alternative or complementary explanation to our stability hypothesis for why eels prefer reclining postures over lunging postures under strong drag.

\section{Supplemental information to Fig. 3}
In this section, we provide supplemental plots to accompany and expand upon the results of Fig.~3 of the main text. 

\subsection{Varying drag}
\begin{figure}[b]
    \centering
    \includegraphics[width=\linewidth]{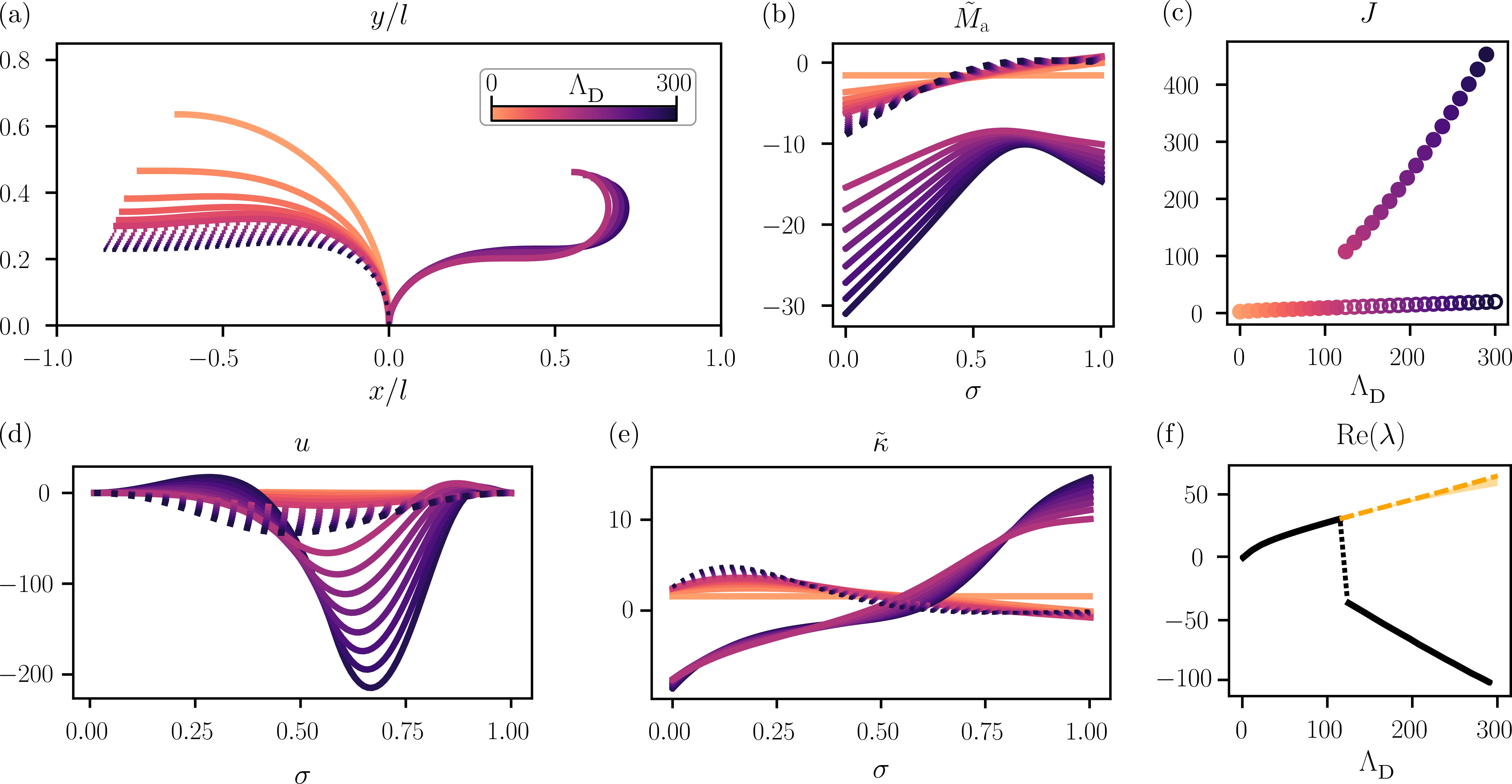}
    \caption{\textbf{Supplemental plots to Fig.~3 for varying $\bm{\Lambda_{\rm D}}$}. Solutions of the optimal control problem in Eq.~(3) for varying values of the drag parameter $\Lambda_{\rm D}$. (a) Body postures for different values of $\Lambda_{\rm D}$. When two solutions exist for a given $\Lambda_{\rm D}$, the forward-leaning and reclining postures are shown as dashed and solid curves, respectively. (b) Active moment profile $\tilde{M}_{\rm{a}}(\sigma)$. (c) Cost $J$, with solid and open circles corresponding to the solid and dashed shapes in (a), respectively. (d) Profiles of the control field $u(\sigma)$. (e) Curvature profile $\tilde{\kappa}(\sigma)$. (f) Real part of the eigenvalues $\lambda$ of Eq.~(4) for the postures in (a), with the dashed yellow and solid black curves corresponding to the dashed and solid solutions, respectively. The dotted portion of the black curve indicates a discontinuity. For all plots we used $\mathcal{K} = 28$, $\mathcal{U} = 600$, $\mathcal{M} = \infty$, $\epsilon = 0.01$.}
    \label{fig:Fig3_supp_SI}
\end{figure}
Here we show additional plots corresponding to Figs. 3(a)--(c), where we fix $\theta(1) = \pi$ and vary the drag parameter $\Lambda_{\rm D}$. We solve the same optimal control problem presented in Eq.~(3) and show the results in Fig.~\ref{fig:Fig3_supp_SI}. Figs.~\ref{fig:Fig3_supp_SI}(a)--(c) correspond to the shape, active moment profile, and cost, respectively, as also shown in Figs.~3(a)--(c) of the main text. In addition, the corresponding control profile $u(\sigma)$ is shown in Fig.~\ref{fig:Fig3_supp_SI}(d) and the curvature profile $\tilde{\kappa}(\sigma)$ in Fig.~\ref{fig:Fig3_supp_SI}(e). In the backward-leaning regime, the minima of $u(\sigma)$ (largest absolute value) correspond to maxima of $\tilde{M}_{\rm{a}}$, where $\tilde{\kappa}(\sigma) = 0$ and the eel is approximately parallel to the current velocity; see Fig.~\ref{fig:Fig3_supp_SI}(a). Finally, in Fig.~\ref{fig:Fig3_supp_SI}(f), we reproduce the inset of Fig.~3(c), showing the real part of the largest eigenvalue obtained from Eq.~(4) of the main text.

\subsection{Varying head orientation}
\begin{figure}[h]
    \centering
    \includegraphics[width=\linewidth]{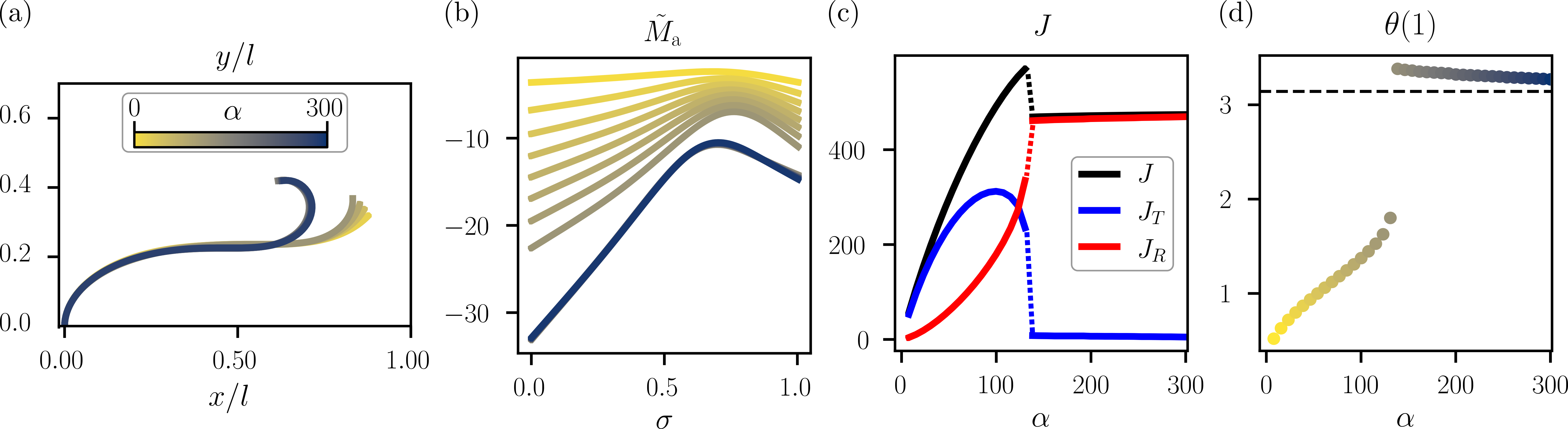}
    \caption{\textbf{Supplemental plots to Fig.~3 for varying $\bm \alpha$}. Optimal solutions to the control problem in Eq.~(3) with $\Lambda_{\rm D} = 300$ and varying $\alpha$. (a) Shapes for different values of $\alpha$. (b) Corresponding active moment $\tilde{M}_{\rm a}$. (c) Total cost, $J(\alpha)$, running cost, $J_R$, and terminal cost, $J_T$, as defined in Eq.~\eqref{eq:cost_decomposition}. The dotted portions indicate discontinuities. (d) Head angle $\theta(\sigma = 1)$ as a function of $\alpha$. For all plots, we used $\mathcal{K} = 28$, $\mathcal{U} = 600$, $\mathcal{M} = \infty$, and $\epsilon = 0.01$.
    }
    \label{fig:control_alpha_SI}
\end{figure}

Now, we fix $\Lambda_{\rm D} = 300$ and vary $\alpha$.
In this case the orientation of the head, $\theta(\sigma = 1)$, is determined by the cost functional $J[u]$, which is decomposed into a running cost $J_R$ and a terminal cost $J_T$:
\begin{equation}
    J[u] = J_{R} + J_T \qquad \text{with} \quad J_R \equiv \int_0^1 \left( \tilde{M}_{\rm a}^2 + \epsilon u^2 \right) {\rm d}\sigma \quad \text{and} \quad J_T \equiv \alpha \big[\theta(1) - \theta^*\big]^2 \;.
\label{eq:cost_decomposition}
\end{equation}
The shapes found from solving the control problem, already shown in Fig.~3(d) of the main text, are reproduced in Fig.~\ref{fig:control_alpha_SI}(a), with the corresponding active moment $\tilde{M}_{\rm a}$ shown in Fig.~\ref{fig:control_alpha_SI}(b). Considering the cost function (see Fig.~\ref{fig:control_alpha_SI}(c)), we find that both the running and terminal costs initially increase with $\alpha$. Around $\alpha^* \approx 130$, there is a jump, with the terminal cost dropping to nearly zero. This corresponds to a discontinuous change in shape where the eel adopts a forward-facing posture; see Figs.~\ref{fig:control_alpha_SI}(a),(d). Interestingly, the total cost is maximal just before the jump and slightly overshoots the cost of forward-facing configurations. This suggests that the state briefly explores an unstable higher-energy solution at intermediate head angles before falling into the lower-energy forward-facing configuration. For this reason, the critical $\alpha^*$ at which we observe a switch in configurations depends not only on the energy landscape but also on the fidelity of the numerical solver and the relative depths of local vs. global energy minima. For larger values of $\alpha$, the cost remains approximately constant, reflecting the fact that $\theta(\sigma = 1) \approx \theta^* = \pi$, cf. Fig.~\ref{fig:control_alpha_SI}(d).

\newpage
\section{Varying bound on active moment}
\begin{figure*}[h]
    \centering
    \includegraphics[width=\linewidth]{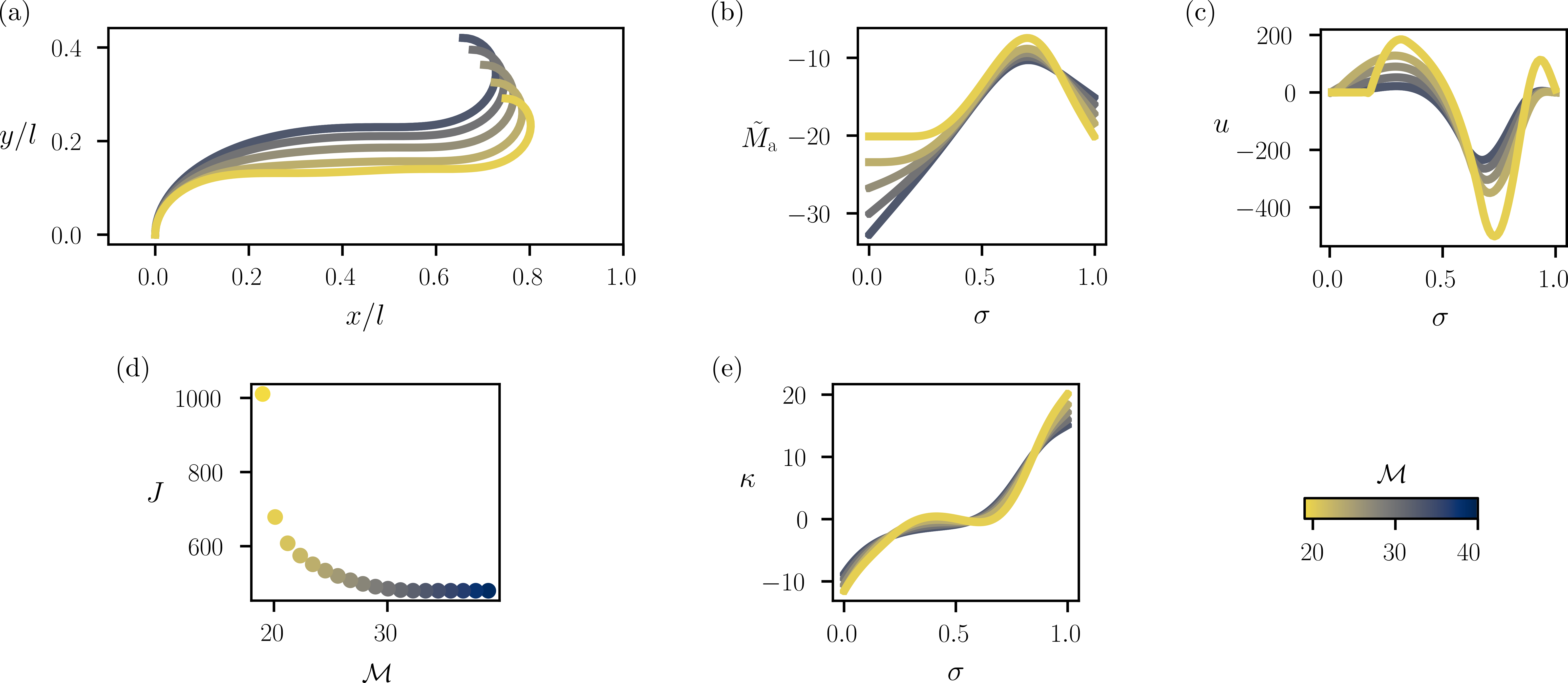}
    \caption{\textbf{Varying bound on active moment}. Solutions to Eq.~(3) of the main text under several values of $\mathcal{M}$, with fixed $\Lambda_{\rm{B}} = 0$ and $\Lambda_{\rm{D}} = 300$. (a) Shapes for different values of $\mathcal{M}$. (b) Corresponding active moment $\tilde{M}_{\rm a}$ and (c) control $u$. (d) Total cost $J(\mathcal{M})$. (e) Curvature $\kappa$.}
    \label{fig:control_bound_M_SI}
\end{figure*}

We examine how the optimal postures in Fig.~3 of the main text change under varying constraints on the maximum muscular activation $\mathcal{M}$. In the main text, we considered the unbounded case, $\mathcal{M} = \infty$. Here, we hold $\Lambda_{\rm B}=0$ and $\Lambda_{\rm D}=300$ fixed and vary $\mathcal{M}$. The results are shown in Fig.~\ref{fig:control_bound_M_SI}. In all cases, the bound on $\mathcal{M}$ is reached at the base of the eel, with $|\tilde{M}_{\mathrm a}(0)| = \mathcal{M}$; see Fig.~\ref{fig:control_bound_M_SI}(b). Tighter bounds on the maximum active moment are compensated by a larger second derivative of $\tilde{M}_{\mathrm a}$ and, therefore, a larger control $u$; see Fig.~\ref{fig:control_bound_M_SI}(c). Given our choice of cost, $j_{\rm R} = \tilde{M}_{\mathrm a}^2 + \epsilon u^2$, the first term decreases under tighter constraints, whereas the second term increases, leading to a rapidly increasing total cost; see Fig.~\ref{fig:control_bound_M_SI}(d). The eel's curvature also increases under tighter constraints, as shown in Fig.~\ref{fig:control_bound_M_SI}(e).

\newpage
\section{Comparison to experiments}
\begin{figure}[h]
    \centering
    \includegraphics[width=\linewidth]{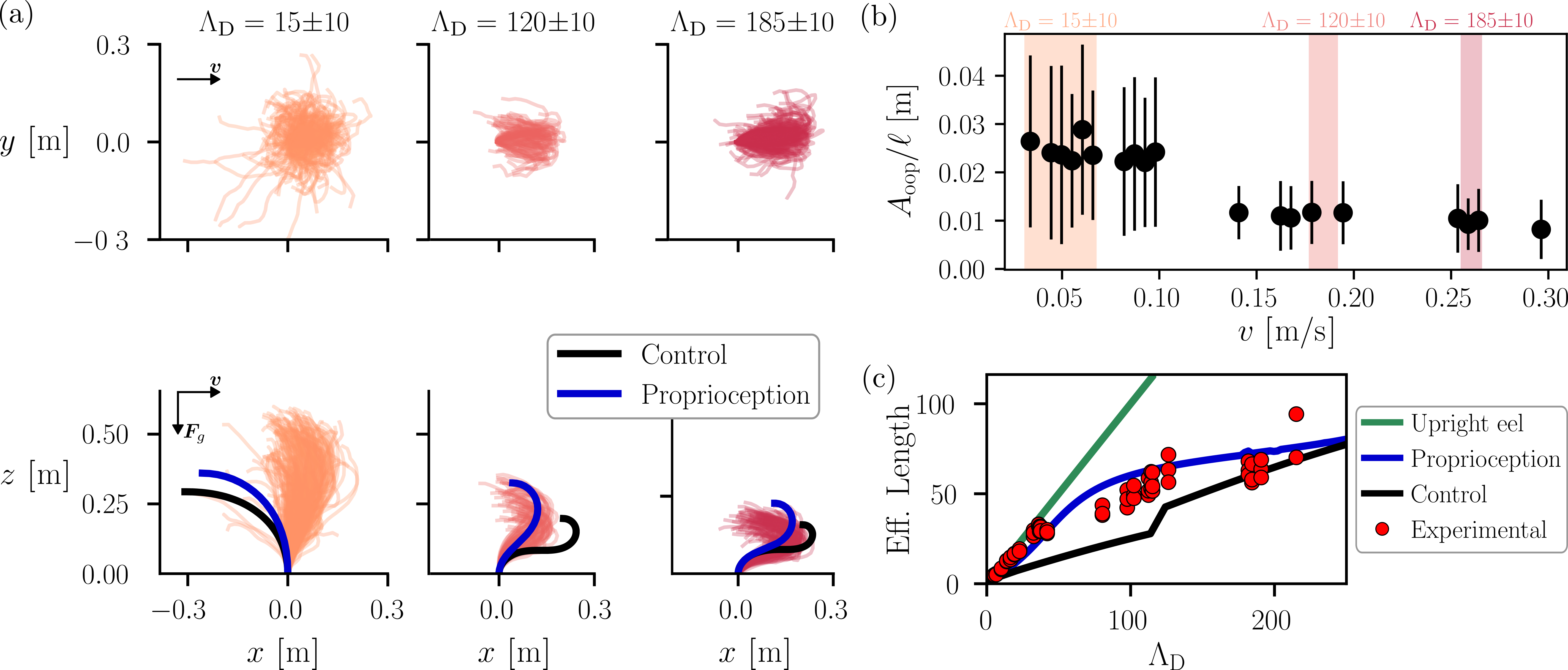}
    \caption{\textbf{Comparison with experiments.} (a) Experimental reconstruction of posture of eels for three different flow velocities, projected into the plane parallel to the seabed (top) and into the plane spanned by current velocity $\bm{v}$ and gravitational force $\bm{F}_g$ (bottom). We overlay the theoretically obtained shapes for the corresponding drag from both the local proprioceptive and the global control approaches. The experimental data are obtained for the following velocity ranges (in m/s): $0.031 < v < 0.068$ (left), $0.177 < v < 0.192$ (center), $0.255 < v < 0.266$ (right). The theoretical curves are obtained for the mean drag values.
    (b) The out-of-plane area $A_{\rm oop}$ is computed as described in the text from reconstructed body curves and evaluated independently for each frame. Points represent bin-averaged values (50 equal-width bins), with error bars indicating standard deviation. The dataset comprises 1,475 frames from 68 distinct eel recordings. The data range corresponding to shapes shown in (a) is indicated by shaded background.
    (c) Effective length $l^* / l$ (see Eq.~\eqref{eq:effective_length}) as a function of drag parameter for an upright eel ($\theta = \pi/2$), optimal eel postures in the proprioception and control models, as well as reconstructed postures of real eels. Each red marker shows the average drag for individual eels averaged over one recording session (see \cite{khrizman2018}). 
    Data reproduced with permission from~\cite{khrizman2018}.
    }
    \label{fig:experimental}
\end{figure}
In this section, we compare in more detail some of our theoretical predictions with data from \cite{khrizman2018} obtained from in vivo recordings of eels in the Red Sea.

In Fig.~\ref{fig:experimental}(a), the reconstructed three-dimensional postures of eels at three different flow velocities are shown~\footnote{Eel postures are estimated using a discrete set of reconstruction points $\{x_i,y_i,z_i\}_{i=1}^{N}$, where $N$ varies from eel to eel (see~\cite{khrizman2018} for detailed methods)}, with the top view ($x-y$-plane, approximately parallel to the seabed) shown in the top panels and the side view ($x-z$-plane, approximately spanned by current velocity $\bm{v}$ and gravitational force $\bm{F}_g$) in the bottom panels. We overlay the shapes obtained from our model, both from using local proprioceptive feedback and global optimal control to model the musculature. For small velocities, the eels are experimentally found to be mostly upright, rather than forward-leaning as predicted by our model. Moreover, their motion is not confined to a single plane, but they frequently move in the direction perpendicular to the flow. This results in a large, roughly isotropic, spread when viewed from the top. Consequently, our model does not capture the experimental observations well at low speeds, where the influence of the current on the shape is presumably weak and the eels instead adopt postures that maximize head height and foraging range, with little regard for drag. 

For stronger currents out-of-plane motion is significantly reduced, and the observed postures lie within a much narrower range of shapes in both the $x-y$-plane and the $x-z$-plane. In this regime, our theoretical predictions agree well with the experimental shapes, particularly for the solution obtained for proprioceptive feedback. 

To justify the two-dimensional assumption that eels remain in the $x-z$-plane spanned by gravity and the current velocity, we quantify the out-of-plane spread observed experimentally. For each eel we compute the out-of-plane area $A_{\rm oop} = \int_0^h |y(z)| dz $ using the trapezoidal rule, and plot the normalized $A_{\rm oop}/l$ as a function of $|\bm{v}|$ in Fig.~\ref{fig:experimental}(b). We find that both the out-of-plane spread and its variance decrease as the eels transition into the reclining posture. 

In Fig.~\ref{fig:experimental}(c) we compare the experimental and theoretical drag forces by comparing the effective length defined in Eq.~\eqref{eq:effective_length}. All three classes of postures yield substantial drag reductions relative to an upright eel. The theoretically obtained curves agree reasonably well with the experimental data.

Regarding the dependence of the exposed length $l$ on the current speed, we found a scaling $l \sim v^{-2/3}$ (see main text). In Ref.~\cite{khrizman2018}, it was reported that $l$ decreased approximately linearly with $v$. Although the limited range of accessible velocities makes it generally difficult to extract precise scaling laws, it is likely that a more complex interplay between posture, exposed length, and overall three-dimensional mobility jointly determines the eels’ behavior.

Finally, Ref.~\cite{ishikawa2022} studied a different species of garden eel, \textit{Heteroconger hassi}, which does not exhibit pronounced postural adaptation, instead reducing its exposed length to reduce drag forces. The authors hypothesized that the species' slender build, and thus reduced musculature compared with \textit{Gorgasia sillneri}, may explain why these eels are unable or unwilling to sustain an aggressive reclining posture. Consistent with this interpretation, Fig.~\ref{fig:control_bound_M_SI} shows that, within our prescribed cost functional, tighter muscular constraints lead to a larger energetic cost. In general, however, detailed data on eel musculature are, to the best of our knowledge, not available, and we are thus restricted to comparing postures rather than underlying mechanical parameters.

\newpage
\section{General posture control}
\begin{figure}[b]
    \centering
    \includegraphics[width=\linewidth]{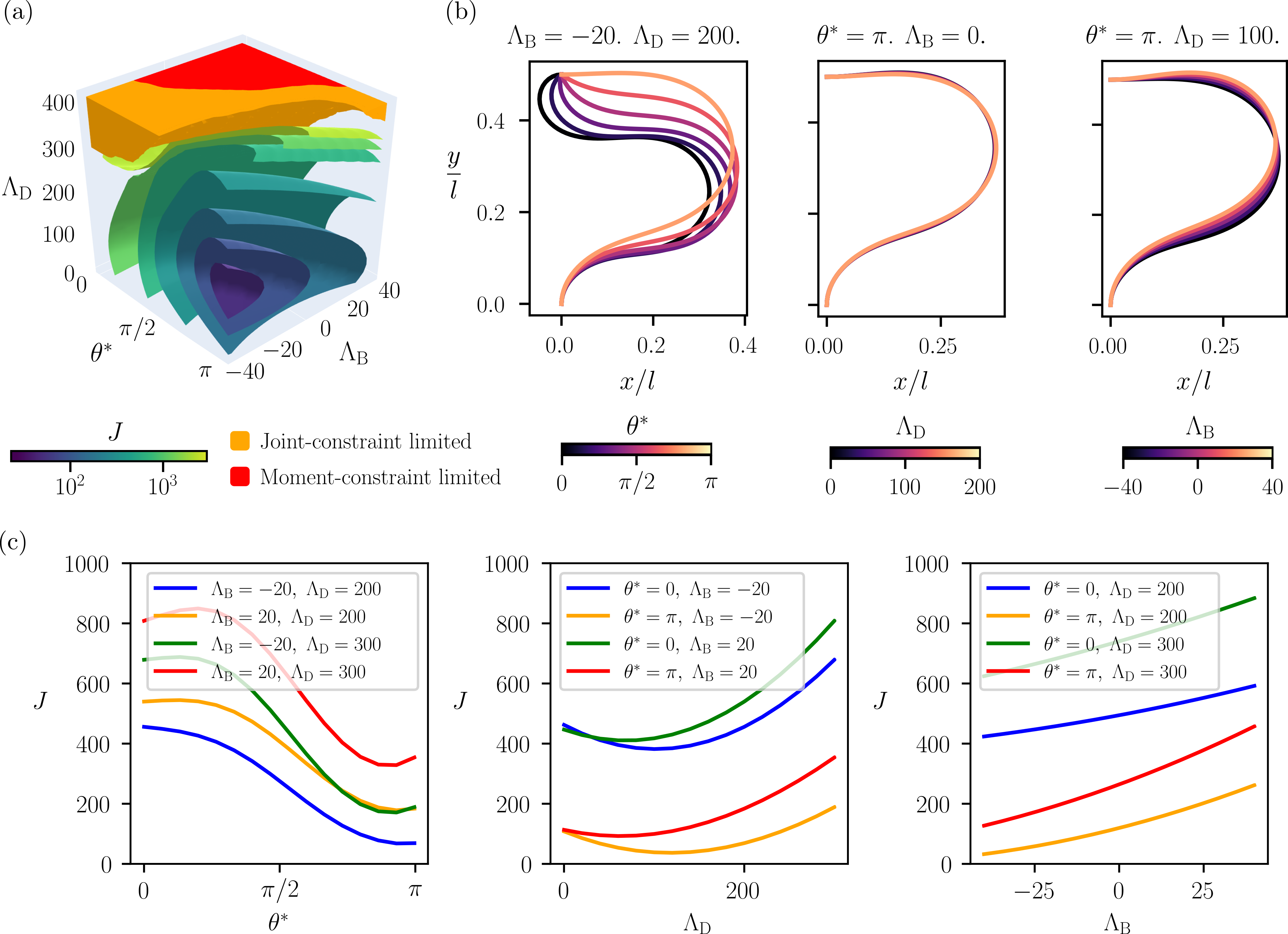}
    \caption{\textbf{General posture control: Example 1.} We solve the general control problem (see Fig.~4 of the main text) with the head fixed at $x(1) = 0$ and $y(1) = 0.5$.  (a) The resulting feasibility diagram is shown. It is identical to that shown in Fig.~4 of the main text. At each point, we solve Eq.~(3) of the main text under four choices of constraints $\mathcal{B} \equiv \{\mathcal{M},\mathcal{K}\}$: unconstrained ($\mathcal{B}_1=\{\infty,\infty\}$), curvature constrained ($\mathcal{B}_2=\{\infty,\mathcal{K}_{\rm b}\}$), active-moment constrained ($\mathcal{B}_3=\{\mathcal{M}_{\rm b},\infty\}$), and jointly constrained ($\mathcal{B}_4=\{\mathcal{M}_{\rm b},\mathcal{K}_{\rm b}\}$). Cost isosurfaces are shown in regions that are feasible under any of the above. The orange region is feasible under $\mathcal{B}_{1-3}$ while the red region is only feasible under $\mathcal{B}_{1,2}$. (b) The shapes along three lines in parameter space are shown on the right. (c) Total cost as a function of one of the three parameters while the other two are kept constant.}
    \label{fig:general_control_1_SI}
\end{figure}

In this section, we provide additional details on the general framework for posture control discussed in the main text. In Fig.~\ref{fig:general_control_1_SI}(a), we reproduce the feasibility diagram from Fig.~4 of the main text. We also show in Fig.~\ref{fig:general_control_1_SI}(b) representative shapes obtained by varying one of the three parameters, $\{\theta^*,\Lambda_{\rm B},\Lambda_{\rm D}\}$, while keeping the other two fixed. Equivalent examples for the cost function are shown in Fig.~\ref{fig:general_control_1_SI}(c). The shape in the case in which $\theta^*$ is varied was already shown in the main text and we find the cost to be nonmonotonic in general. We find that varying $\Lambda_{\rm D}$ produces only minimal changes in shape but the total cost $J$ is nonmonotonic and increases significantly for large $\Lambda_{\rm D}$. In particular, for sufficiently large $\Lambda_{\rm D}$ the shape is always infeasible in the presence of constraints on the active moment. The entire parameter space considered is accessible if only curvature is constrained. In contrast, varying $\Lambda_{\rm B}$ produces small changes in shape while the cost $J$ varies monotonically and over a much smaller range, see Fig.~\ref{fig:general_control_1_SI}~(c).

\begin{figure}
    \centering
    \includegraphics[width=\linewidth]{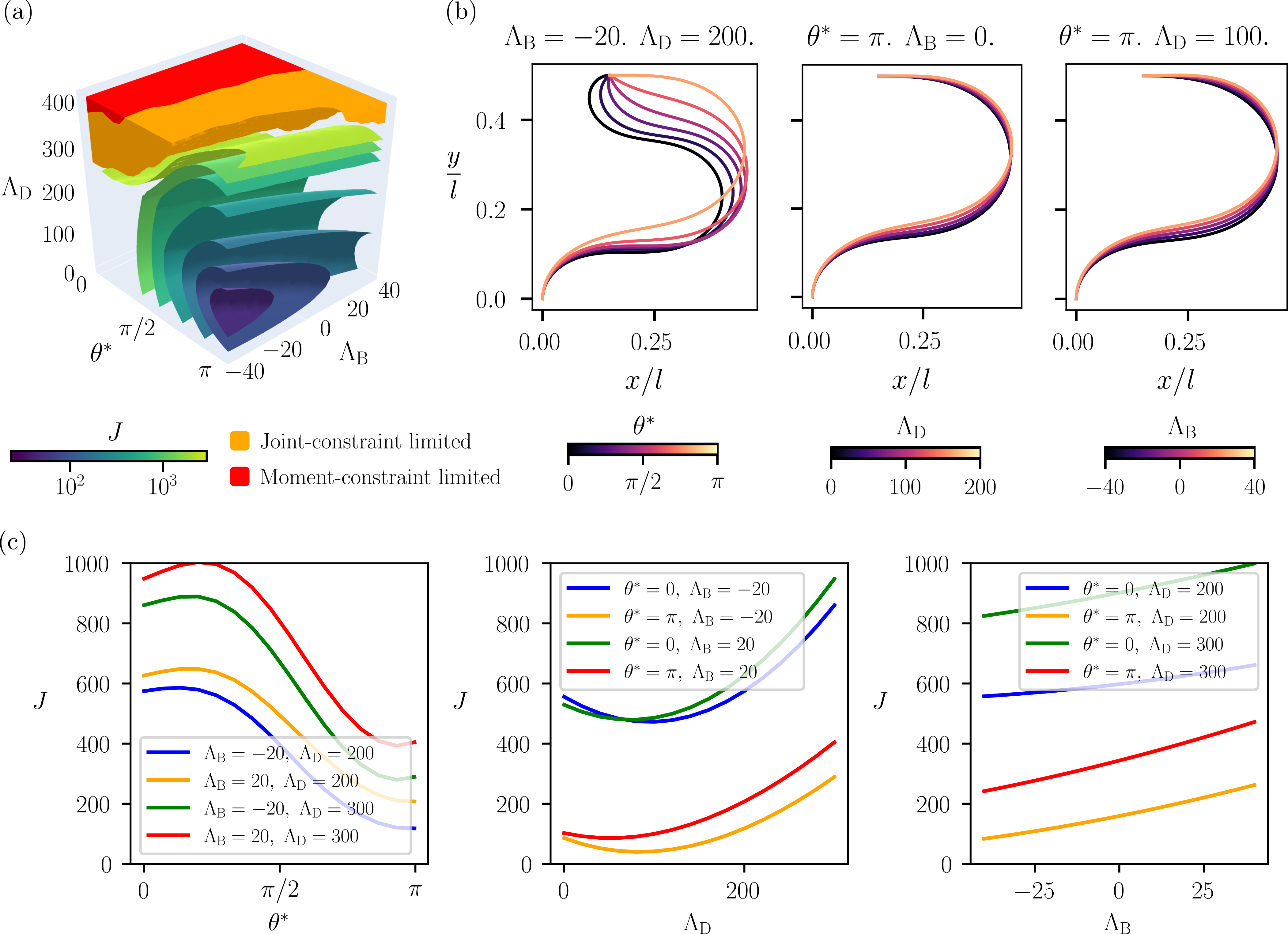}
    \caption{\textbf{General posture control: Example 2.} Results equivalent to those in Fig.~\ref{fig:general_control_1_SI}, but for $x(1) = 0.15$ and $y(1) = 0.5$.}
    \label{fig:general_control_2_SI}
\end{figure}

In Fig.~\ref{fig:general_control_2_SI}, we show analogous data for the case in which the head position is prescribed as $x(1) = 0.15$ and $y(1) = 0.5$. The results are qualitatively similar, with the main difference being the change in the red region.

\newpage
\section{Numerical methods}
For both the proprioception model and the passive filament, we solve the associated boundary value problems numerically using a custom \texttt{Python} script based on the \texttt{scipy.integrate.solve\_bvp} function. To solve the optimal control problem, we formulate the optimal-control problem using \texttt{CasADi}~\cite{andersson2019} and solve the resulting nonlinear program with \texttt{Ipopt}. The problem is discretized using 100 intervals with degree-3 Legendre collocation. The code and implementation details are available at \url{https://github.com/apearl16/Posture_selection_active_filaments}.

\bibliographystyle{apsrev4-2}
\bibliography{Biblio.bib}